\documentstyle[11pt,newpasp,twoside]{article}
\def\edcomment#1{\iffalse\marginpar{\raggedright\sl#1\/}\else\relax\fi}
\reversemarginpar

\begin{document}

\title{Light curves and metal abundances of RR Lyrae variables in the bar of
 the Large Magellanic Cloud}

\author{Clementini G., Bragaglia A., Di Fabrizio L.}
\affil{Osservatorio Astronomico di Bologna, Via Ranzani 1, I-40127 Bologna, 
Italy}
\author{Carretta E., Gratton R.G.}
\affil{Osservatorio Astronomico di Padova, Vicolo dell'Osservatorio 5, 
I-35122 Padova, 
Italy} 

\begin{abstract}
The Large Magellanic Cloud (LMC) is widely considered a corner-stone of 
the astronomical distance scale, however
a difference of 0.2--0.3 mag exists in its 
distance as predicted by the {\it short} and {\it long} 
distance scales. 
Distances to the LMC from Population II objects are founded on the 
RR Lyrae variables.
We have undertaken an observational campaign devoted to  
the definition of the average apparent luminosity and to the study of 
   the mass-metallicity relation for RR Lyraes in the
   bar of the LMC. These are compared with analogous quantities 
for cluster RR Lyraes. The purpose is to see whether an intrinsic 
   difference in luminosity,
   possibly due to a difference in mass, might exist between field  and 
cluster RR Lyraes, which could be 
   responsible for the well-known dichotomy between {\it short} and {\it long} 
   distance scales.
Preliminary results  are presented on the V and B$-$V light curves, the
average apparent visual magnitude, and the pulsational properties of 102
 RR Lyrae in the bar of the LMC,
observed at ESO in January 1999. The photometric 
data are accurately tied to the Johnson photometric
system. Comparison is presented with the photometry of RR Lyraes in the 
bar of the LMC obtained by the MACHO collaboration (Alcock et al. 1996).
Our sample includes 9 double-mode RR Lyraes selected from Alcock et al. 
(1997) 
for which an estimate of the metal abundance from 
the $\Delta$S method is presented.
\end{abstract}


\section{Observations, data reduction and calibration}
We have collected B,V CCD photometry 
   in two 13$^{\prime}$ x 13$^{\prime}$ fields located  close 
   to the bar of the LMC and overlapping with fields \#6 and \#13 of the MACHO 
   microlensig experiment (see http://wwwmacho.mcmaster.ca), 
using the 1.5m Danish telescope in La Silla. 
The photometric data set consists of 58 V and 25 B frames of each field.
118 variables were identified in the two fields (62 
{\it ab} type RR Lyrae, 30 RR$_{c}$, 10 RR$_d$, 6 Cepheids, 9 eclipsing 
binaries and
1 $\delta$ Scuti). Photometry was accurately tied to the Johnson standard 
photometric system using a large number of standard stars from Landolt (1992). 

Low resolution spectra (R=450, res.element=9 \AA) were obtained 
   for 7 of the RR$_d$ variables with EFOSC2 at the 3.6 m ESO telescope, 
and metal abundances have been derived using 
the $\Delta$S technique (Preston 1959). For calibration purposes we took also spectra 
at minimum light of 8 field RR Lyraes of known $\Delta$S (of which
a {\it c} type followed along the pulsation cycle), and 
of 14 stars of the open cluster 
Collinder 140, which contains spectral type 
standard stars.

Photometric data were analyzed using the package DoPHOT (Schechter, Mateo, 
\& Saha, 1993).                
Spectroscopic data were reduced using the standard IRAF packages for long
slit spectra. 
Total numbers of 28000 and 25000, and 23000 and 19000 objects were measured 
in the V and B frames of the two fields, respectively. 
\begin{figure}
\vspace{6.1cm}
\includegraphics{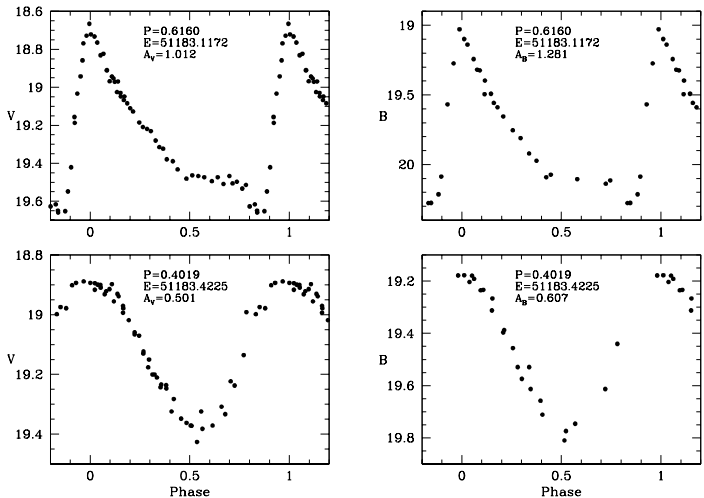}
\caption{V and B light curves of RR Lyrae variables in our sample}
\end{figure}  
The average magnitude 
the LMC clump stars is $<V>$ = 19.202 ($\sigma$=0.202; 8979 stars)
The comparison with the Alcock et al. 
(1997, hereinafter A97) $<V>$ for the clump, as can be read from their Fig. 3,  
shows that our value is about 0.10 mag "fainter".

\section{Identification, period search and pulsational properties of the 
RR Lyrae variables}
Variables were identified on the V and B frames independently.
 Periods were defined using 
the program GRATIS (GRaphycal Analyzer TIme Series; Montegriffo, Clementini, 
\& Di Fabrizio 1999, 
in preparation) 
 which was run on the differential photometry of the variables
   with respect to a number of stable reference stars.
The period search procedure was to perform a Lomb analysis
 (Lomb 1976) on a wide period interval first, and then use a Fourier 
analysis to refine periods and find the best fitting models.
   Average residuals from the best fitting models for single-mode 
pulsators with well
   sampled light curves are 0.02--0.03 mag in V, and 0.04--0.06 mag in B. 
Figure 1 shows examples of the V and B light curves of an {\it ab}, a 
{\it c} type RR Lyrae in our sample. 
The period distribution of the 
{\it c} type RR Lyraes in our fields peaks at  $<$P(RR$_c$)$>$=0.314 $\pm$ 0.047 
(average on 30 stars), while $<$P(RR$_{ab}$)$>$=0.577$\pm$0.077 
(average on 60 stars) to compare with 0.342 and 0.583 of Alcock et al. (1996, 
hereinafter A96).
Our shortest period {\it ab} type RR Lyrae has period 0.318 d,
 and there are two other RR$_{ab}$'s with periods around 0.40 d.
We derived  $<V>$ and $<B>$ intensity average magnitudes as well as  V and B 
amplitudes (A$_V$ and A$_B$) for all variables with complete light curves. 
The average $<V>$ apparent magnitude of the RR Lyraes in our sample 
is : $<V>=19.325 \pm 0.170$  (75 stars), to compare with $<V>$=19.4  from 
A96. On the assumption that :  E(B$-$V)=0.10 (Bessel 1991) and 
A$_V$=3.1[E(B$-$V)] for the LMC 
we find :
\medskip
\par\noindent
$<V_0>$=19.015$\pm0.020$ at [Fe/H]$\sim-1.5$~~{\it field} RR Lyraes, this paper
\par\noindent
$<V_0>$=19.09   ~~~~~~~~~~ at [Fe/H]$\sim-1.7$~{\it field} RR Lyraes, A96
\par\noindent 
$<V_0>$=19.06$\pm0.06$~~~~~~~~~~~~~~~~~~~~~~~~~~~{\it field} RR Lyraes, 
Kinman et al. (
1991)  
\par\noindent
$<V_0>$=18.94$\pm0.040$~~ at [Fe/H]=$-$1.9~~{\it cluster} RR Lyraes, Walker (1992)
\medskip
\par\noindent
Allowing for the 0.4 dex difference in [Fe/H] our $<V_0>$ is in very
good agreement with Walker (1992), thus showing that there is no clear 
evidence for a difference in luminosity between field and cluster RR Lyraes in the 
LMC. 

\section{Spectroscopy of the double-mode RR Lyraes}
  Spectra for 7 of the RR$_d$ variables falling in our fields were obtained
  at phases corresponding to the minimum light.
Metallicities were inferred from these spectra using the $\Delta S$ index 
 and 
the Clementini et al. (1995) calibration of 
$\Delta$S in terms of metallicity ([Fe/H]=$-0.194\times\Delta$S $-$0.08). 
\begin{table*}
\begin{center}
\scriptsize
\caption{$\Delta$S values, metal abundance and masses of RR$_d$ pulsators}
\begin{tabular}{cllc}
\hline\hline
\multicolumn{1}{r}{N(A97)}&
\multicolumn{1}{l}{~$\Delta$S}&
\multicolumn{1}{l}{[Fe/H]}&
\multicolumn{1}{c}{~M/M$_{\odot}$}\\
\hline
~2&~8.6&$-$1.74&0.60\\
~5&11.3:&$-$2.28:&0.61\\
45&~7.9&$-$1.62&0.65\\
48&~5.2&$-$1.09&0.67\\
49&~7.3&$-$1.50&0.69\\
61&~5.9&$-$1.23&0.71\\
67&~8.8&$-$1.78&0.81\\
\hline
\end{tabular}
\end{center}
\end{table*}
Dealing with variables which pulsate both in the fundamental and first
overtone, the question arises whether these 
stars should be treated as {\it ab} or {\it c} type pulsators 
in measuring $\Delta$S. Following Kemper (1982) and discussions in Clement,
 Kinman, \& Suntzeff (1991),  we considered our targets as 
{\it c} type variables. 
$\Delta$S values were thus measured from spectra with Hydrogen spectral type later 
than  A8, and applying phase corrections derived from the field RR$_c$ T Sex.
Table 1 lists the  
$\Delta$S values and corresponding metallicities we derived for our
targets. 
  Values for star \#5 are rather uncertain since the spectrum of this 
  star has very low S/N. Errors on the quoted $\Delta$S are of the order of
0.7--1 $\Delta$S subclasses, corresponding to an error of about 0.20--0.30 dex 
in [Fe/H].

\section{The mass-metallicity relation for the RR Lyrae stars}

  A97 provides P$_0$/P$_1$ ratios for the 7 double-mode
  variables in Table 1. These ratios can be used together with 
  Petersen diagrams (Petersen 1973) and 
  Bono et al. (1996) loci of model pulsators
  to estimate the masses of our targets (see e.g. Figure 2 of A97). Masses 
  obtained with this procedure are listed in the last column of Table 1 and 
  plotted against metallicity in Figure 2. Also shown is the mass-metallicity
  relation defined by double-mode pulsators 
in the globular clusters M68 (Walker 1994), 
  M15 (Nemec 1985)  and
  IC 4499 (Clement et al. 1986, Walker \& Nemec 1996) and two  RR$_d$'s 
  in the Milky Way (Clement et al., 1991). Although there is some scatter, and there are only few field objects,
  most of the LMC RR$_d$'s seem to
  follow the general mass-metallicity relation defined by the cluster 
  RR$_d$'s. Hence, no clear-cut evidence is found
  for a difference in mass between field and cluster RR Lyraes.
\begin{figure}
\vspace{6cm}
\includegraphics{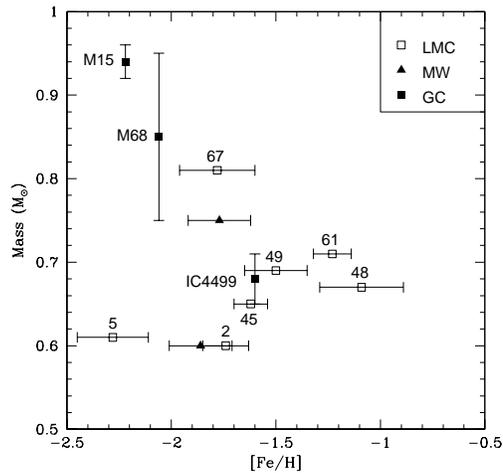}
\caption{Mass-metallicity relation of field and cluster double-mode pulsators}
\end{figure}  

\end{document}